\documentclass[5p, times]{elsarticle}

\usepackage{hyperref}

\usepackage[utf8]{inputenc}  
\usepackage[T1]{fontenc} 
\usepackage{amsmath}
\usepackage{amssymb}
\usepackage{amsfonts}
\usepackage{amsbsy}
\usepackage{graphicx}
\usepackage{color}

\usepackage{tabularx}

\journal{Comptes Rendus de l'Académie des Sciences}

\begin{document}

\begin{frontmatter}


\title{State of the art in the determination of the fine structure constant and the ratio $h/m_\mathrm{u}$.\\
L'état de l'art de la détermination de la constante de structure fine et du rapport $h/m_\mathrm{u}$.
}


\author[lkbjussieu]{Pierre Cladé}
\author[lkbjussieu]{François Nez}
\author[lkbjussieu]{François Biraben}
\author[lkbjussieu, cnam]{Saïda Guellati-Khelifa}

\address[lkbjussieu]{Laboratoire Kastler Brossel, Sorbonne Universit\'e, CNRS, ENS-PSL University, Coll\`ege de France, 4 place Jussieu, 75005 Paris}
\address[cnam]{Conservatoire National des Arts et M\'etiers, 292 rue Saint Martin, 75003 Paris, France}

\begin{abstract}
The fine structure constant $\alpha$ and the ratio $h/m_{\mathrm{u}}$ between the Planck constant and the unified atomic mass are keystone constants for the determination of other fundamental physical constants, especially the ones involved in the framework of the future International System of units.
This paper presents how these two constants, which can be deduced from one another, are measured. We will present in detail the measurement of $h/m_\mathrm{Rb}$ performed by atomic interferometry at the Laboratoire Kastler Brossel in Paris. 
This type of measurement also allows a test of the standard model to be carried out with unparalleled accuracy. 
\medskip\\
\textbf{Résumé}\\
La constante de structure fine $\alpha$ et le rapport $h/m_{\mathrm{u}}$ entre la constante de Planck et la masse atomique unifiée sont des constantes clés pour la détermination d'autres constantes physiques fondamentales, notamment celles impliquées dans le futur système international d'unités.
Cet article présente comment ces deux constantes, qui peuvent être déduites l'une de l'autre, sont mesurées. Nous présenterons en détail la mesure de $h/m_\mathrm{Rb}$ effectuée par interférométrie atomique au Laboratoire Kastler Brossel à Paris. 
Ce type de mesure permet également d'effectuer un test du modèle standard avec une précision inégalée. 

\end{abstract}

\begin{keyword}
Fine structure constant; electron moment anomaly; atom interferometry; international system of units.\\
\textit{Mots clés:} Constante de structure fine, moment magnétique anomal de l'électron, interférométrie atomique, système d'unité international.
\end{keyword}

\end{frontmatter}


\section{Introduction}

\subsection{Determinations of $\alpha$ and $h/m_\mathrm{u}$}

Since its discovery at the beginning of the 20$^\mathrm{th}$ century, the fine structure constant $\alpha$ remains one of the most fascinating fundamental constants, as it is dimensionless. Currently it plays not only a central role in the Physics of the 21$^\mathrm{st}$ century as it allows to test the most accurate theories such as quantum electrodynamics (QED) \cite{Bouchendira2011, Hanneke2008, Aoyama2012, Parker2018} and the stability of fundamental constants ($\dot{\alpha}/\alpha$) (see for example review by J.P. Uzan \cite{Uzan}) but also, in a practical way, in the proposed redefinition of the international system of units (SI) \cite{Mills2011}.

The name of the fine structure constant originates from the Sommerfeld model \cite{Sommerfeld1916}. It was intended to explain the fine structure of the hydrogen spectral lines, unaccounted for in the Bohr model. The Sommerfeld model combines the theory of relativity with the Bohr model. The constant $\alpha$ appears in the velocity of the electron ($v_\mathrm{e}$) on its first orbit around the proton ($v_\mathrm{e}=\alpha\times c$, where $c$ is the velocity of light). The expression for $\alpha$ is:
\begin{equation}
\alpha=\frac{e^2}{4 \pi \epsilon_{0} \hbar c}\label{eqalpha}
\end{equation}
where $e$ is the charge of the electron, $\epsilon_{0}$ the vacuum permittivity and $\hbar=h/2\pi$ in which $h$ is the Planck constant.


The modern understanding of $\alpha$ is that it sets the scale of the electromagnetic interaction. Consequently, many experiments in which a charged particle interacts with an electromagnetic field can be used to determine $\alpha$. In  1998, the experiments considered by the CODATA task group on fundamental constants to give the best estimate of the fine structure constant value ranged from solid state physics and atomic physics to quantum electrodynamics \cite{CODATA98}. Currently, the most acurate determination of the fine structure constant comes mainly from two methods. 

\begin{figure}
\includegraphics[width=.9\linewidth]{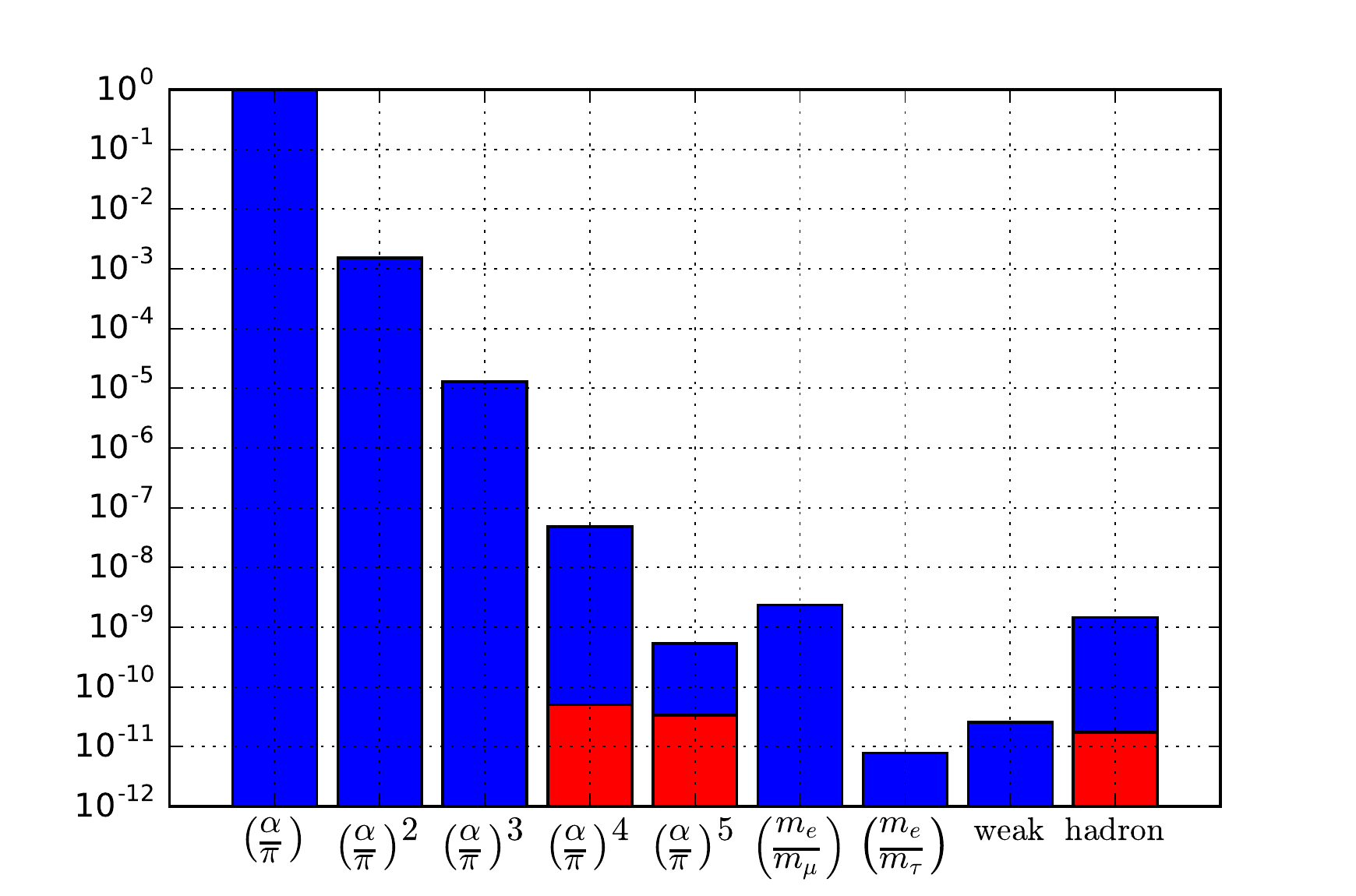}
\caption{\label{fig:calcul_ae} Relative contributions to the electron anomaly of the different terms of equation \ref{eq:ae}, in red their uncertainties.}
\end{figure}

The first method combines the measurement of the electron's gyromagnetic anomaly, ${a_\mathrm{e}}$, with the QED calculations. Indeed, it is possible to extract the value $\alpha$ from the following equation: 
\begin{multline}
{a_\mathrm{e}} = A_1\left(\frac{\alpha}{\pi}\right)
+ A_2\left(\frac{\alpha}{\pi}\right)^2
+ A_3\left(\frac{\alpha}{\pi}\right)^3
+ A_4\left(\frac{\alpha}{\pi}\right)^4 \\
+ A_5\left(\frac{\alpha}{\pi}\right)^5
+\ldots + {a_\mathrm{e}}\left(\frac{{m_\mathrm{e}}}{m_\mu}\right) 
 +  {a_\mathrm{e}}\left(\frac{{m_\mathrm{e}}}{m_\tau}\right) \\
+ {a_\mathrm{e}}(\mathrm{weak}) + {a_\mathrm{e}}(\mathrm{had})
\label{eq:ae}
\end{multline}
Figure \ref{fig:calcul_ae} represents the relative amplitude of all the terms of this equation with their respective uncertainty. The coefficients $A_i$ are dimensionless numbers calculated using Feynmann diagrams. The first ones ($A_1$, $A_2$ and $A_3$) are known analytically. The coefficients $A_4$ and $A_5$ are calculated numerically, the last one involving more than 10 000 diagrams \cite{Aoyama2012, LAPORTA2017232, PhysRevD.97.036001}. Small contributions come from different effects: the contributions involving other leptons in the loop (muons and taus), that scale with the mass ratios and the contribution due to weak and hadronic interactions. The uncertainty of all the terms are estimated to be below $10^{-10}$, which means that, at this accuracy, a measurement of ${a_\mathrm{e}}$ is equivalent to a measurement of $\alpha$. 

The second method involves the measurement of the ratio between the Planck constant $h$ and the mass of an atom $m_\mathrm{At}$. The fine structure constant is measured via the relation 
\begin{equation}
\alpha^2 = \frac{2R_\infty}c \frac{m_\mathrm{At}}{{m_\mathrm{e}}}\frac h{m_\mathrm{At}}.
\end{equation}
The Rydberg constant $R_\infty$ is known to $6\times 10^{-12}$ and the atom to electron mass ratio $\frac{m_\mathrm{At}}{{m_\mathrm{e}}}$ to better than $10^{-10}$ for many atoms. In this equation, the measurement of the ratio $\frac h{m_\mathrm{At}}$ limits the determination of $\alpha$. In order to link back the measurements of $\frac h{m_\mathrm{At}}$ made on different atoms to a single fundamental constant, we introduce the unified atomic mass $m_\mathrm{u} = m_\mathrm{^{12}C}/12$ and the relative atomic mass of an atom ($A_r(\mathrm{At}) = m_\mathrm{At}/m_\mathrm{u}$) and calculate the ratio $h/m_\mathrm{u} = A_r(\mathrm{At}) \frac h{m_\mathrm{At}}$.

\begin{figure}
\includegraphics[width=.95\linewidth]{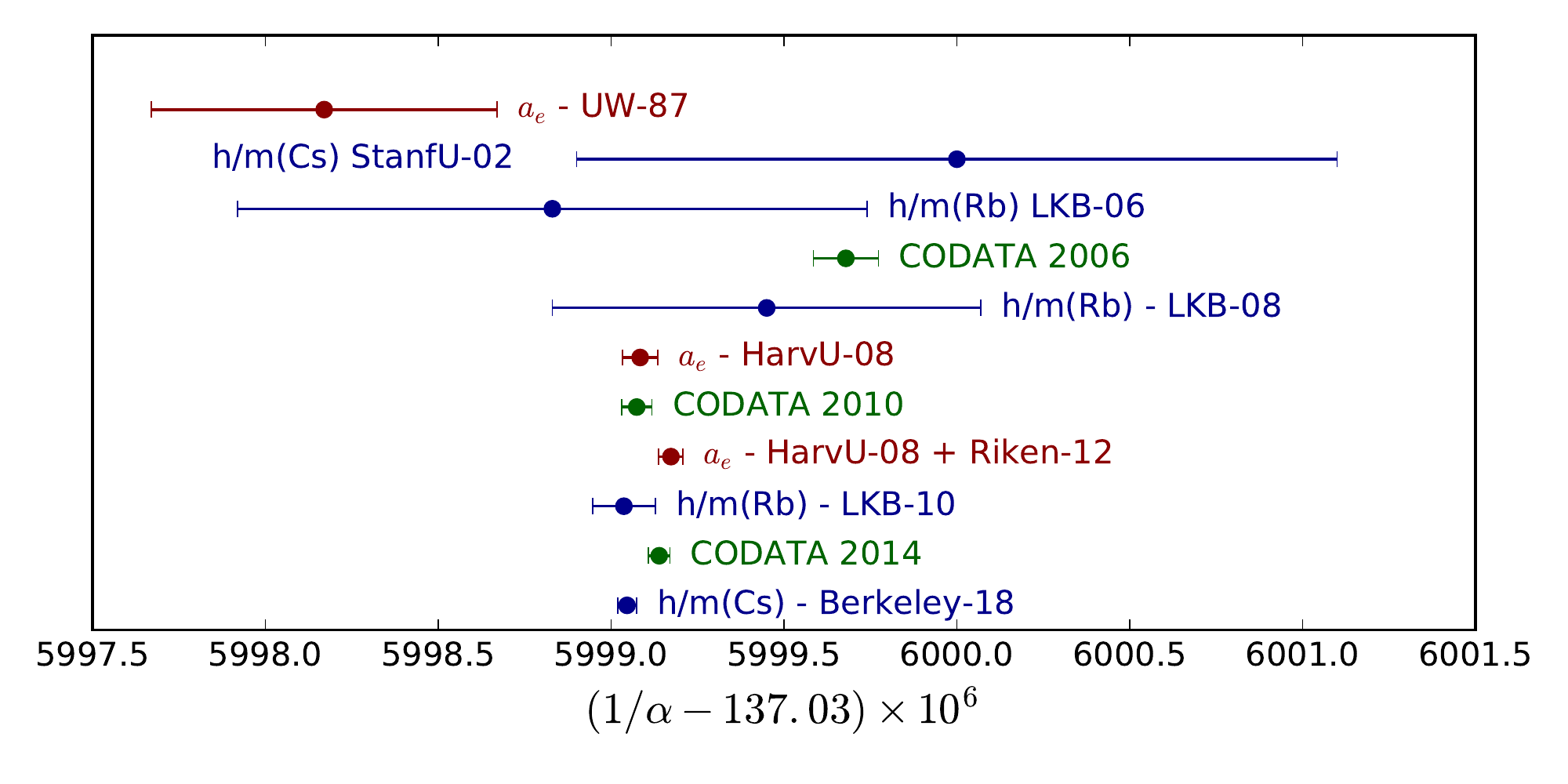}
\caption{\label{fig:mesures} Comparison of different determinations of $\alpha$ with uncertainties below $10^{-8}$. In red, determinations from the magnetic moment of the electron, in blue from $h/m_{\mathrm{At}}$ and in green from the CODATA. UW: University of Washington \cite{VanDick}; LKB: Laboratoire Kastler Brossel (Paris) \cite{clade:033001, Cadoret2008, Bouchendira2011}; Berkeley \cite{Parker2018} ;HarvU: Harvard University \cite{Hanneke2008}; Riken \cite{Aoyama2012}; CODATA \cite{CODATA2006, CODATA12, CODATA2014} }
\end{figure}

On Fig.~\ref{fig:mesures} we have plotted different determinations of $\alpha$ with uncertainties below $10^{-8}$. 
The first one was deduced from the electron anomaly in 1987, by the group of Dehmelt at university of Washington \cite{VanDick}. Twenty years later a new measurement was obtained by the group of Gabrielse at Harvard University \cite{Hanneke2008}. Determination of $\alpha$ using measurement of $h/m$ using atom interferometry was introduced by the group of S. Chu, who published a value in 2002 \cite{Wicht:02}. Our group published three values of $h/m_\mathrm{Rb}$ \cite{clade:033001, Cadoret2008, Bouchendira2011}, the last one with an uncertainty of $6.6\times 10^{-10}$. In 2018, the group of H. Müller, who has taken over S. Chu experiment, published a value $\alpha$ with an uncertainty of $2\times 10^{-10}$ \cite{Parker2018}. This is the first time, since the measurement of Dehmelt in 1987, that the most precise determination of $\alpha$ is obtained from a different method than the determination of the electron magnetic moment. 

The CODATA Task Group on Fundamental Physical Constants publishes every four years a set of recommend values of fundamental constants obtained from all available measurements. The existence of two methods to determine $\alpha$ is important, because it increases the reliability of the adjustment. For example, in the past, the value of $\alpha$ changed significantly due to an error discovered in the calculation of the coefficient $A_4$ \cite{Aoyama2007}.

Furthermore, comparing the two values allows to perform a test of the underlying physics, in this case the equation \ref{eq:ae}. Experiments are so precise that they allow to test QED at the tenth order and observe on lab-size experiments the correction due to muons (\cite{Bouchendira2011}). The recent determination obtained at Berkeley is slightly shifted from the ${a_\mathrm{e}}$ measurement by about 2.5~$\sigma$. Authors claim that this tension rejects dark photons as the reason for the unexplained part of the muons's magnetic moment at a 99\% confidence level \cite{Parker2018}. The $2.5~\sigma$ discrepancy is insufficient to conclude that there are new terms in ${a_\mathrm{e}}$ coming from new particles but it may be a sign of physics beyond the standard model.  This warrants further investigation.

\subsection{The role of $\alpha$ and $h/m_\mathrm{u}$ in the SI}

The new system of units that will be implemented in 2019 is build on fundamental constants. Among them, the fine structure constant plays an important role for the electrical units, and the ratio $h/m_\mathrm{u}$ for the mass unit. For both constants, we will distinguish between the role they play in the CODATA for fixing the fundamental constants used for the new definition and the role they will play in the new system. 

\begin{figure}
\includegraphics[width=.95\linewidth]{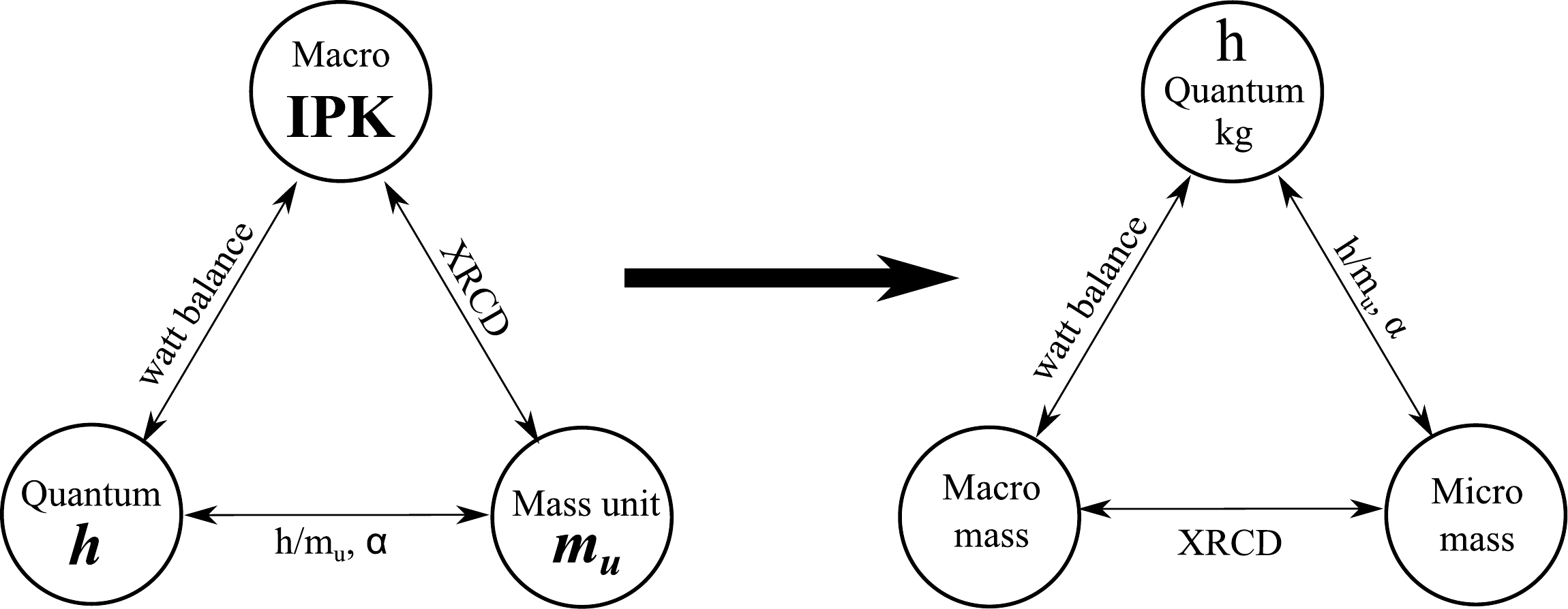}
\caption{\label{fig:triangle} 
In the current SI, the ratio $h/m_u$ provides a direct comparison between
the watt balance experiment and the XRCD experiment. In the future SI this ratio
would achieve the \textit{Mise en pratique} of the kilogram at the atomic scale and at the
macroscopic scale using the XRCD method.}
\end{figure}

In the new SI, the Planck constant $h$ will have a fixed value. This value has been chosen using the adjustment made by the CODATA. Two kinds of experiments were involved. The first one, the watt balance (or Kibble balance), measures the ratio $h/M$ between the Planck constant and a macroscopic standard mass $M$ \cite{Kibble1990, Steiner2007,Williams1998}. In the previous SI, it leads to a determination of $h$; in the new SI, it gives a direct measurement of a macroscopic mass. The second experiment is the x-ray-crystal-density (XRCD) method \cite{Andreas2011}, which directly measures the ratio $M/m_\mathrm{Si}$ between a macroscopic mass (the mass of a silicon sphere) and an atomic mass. This is done by counting the number of atoms in the silicon sphere. In the previous SI, it gives a determination of $m_\mathrm{Si}$ and therefore of $m_\mathrm{u}$ or the Avogadro constant (again we assume that $A_r(\mathrm{Si})$ is well known). The ratio $h/m_\mathrm{u}$ is used to provide a determination of $h$ from this measurement and to compare it with the watt balance measurement (see Fig.~\ref{fig:triangle}). An uncertainty better by one order of magnitude than the uncertainty involved in watt balance or XRCD method was important for the new definition, as it allows to consider both experiments equivalent in the adjustment of fundamental constants.

In the new SI where $h$ will be fixed, the ratio $h/m_\mathrm{u}$ will be a measurement of $m_\mathrm{u}$ and therefore will be the way to link the atomic mass units to the SI. Furthermore, it will be used for the \textit{mise en pratique} of the kilogram with the XRCD method \cite{CCMRQ2014, Clade_metrologia_2016, Fujii2016}. 

We emphasize that in the previous SI, the ratio $h/m_\mathrm{u}$ and the so-called Avogadro Planck constant ($N_{\mathrm{A}}h$) were equivalent. They are indeed related trough the followin-
\begin{equation}
N_{\mathrm{A}}h=\frac {h}{m_\mathrm{u}}\frac {M(^{12}\mathrm{C})}{12}
\label{eq:NAh}
\end{equation}
where $M(^{12}C)$ is carbon molar mass which is equal exactly to $12\times 10^{-3}\mathrm{kg/mol}$ in the previous SI. In the new SI, the Avogadro constant $N_{\mathrm{A}}$, which is used by chemists to quantify and identify an amount of substance with atoms and molecules, is fixed. This breaks the link between atomic masses and molar masses. Consequently $M(^{12}$C) will no longer be exactly 12 g/mol, but will be determined from equation \ref{eq:NAh} using the ratio $h/m_{\mathrm{u}}$.

The new system of unit redefines also the electrical units. This is done by fixing the value of the elementary charge $e$. The value of the Josephson constant ($2e/h$) and of the von Klitzing constant ($h/e^2$) are therefore fixed and measurements of voltage or current are reduced to frequency measurements.  

In the definition of the fine structure constant 
\begin{equation}
\alpha = \frac{e^2}{4\pi \epsilon_0 \hbar c}
\label{eq:alpha}
\end{equation}
within the framework of the previous SI, the vacuum permittivity $\epsilon_0$, and the speed of light $c$ are fixed. Because $\alpha$ is known with a much smaller accuracy than the one of $h$ and $e$, those two constants are equivalent in regard to the CODATA adjustment. Actually, there are no direct measurement of $e$ with sufficient accuracy, and the value used for the new SI was thus mainly obtained using equation \ref{eq:alpha} and the value of $h$ determined as described above. In the new SI, as $h$ and $e$ are fixed, a measurement of $\alpha$ will provide a measurement of $\epsilon_0$ (or $\mu_0$). 

\bigskip

The next section of this paper will be devoted to the experiment we are conducting in Paris to measure the ratio $h/m_{Rb}$. Our last value was published in 2011 and remained the most precise direct measurement of $h/m_\mathrm{u}$ until the recent measurement performed at Berkeley \cite{Parker2018}. Both experiment are based on atom interferometry and use similar techniques. 

\section{Determination of the ratio $h/m_{\mathrm{Rb}}$}
\subsection{Principle}
The ratio $h/m_{\mathrm{Rb}}$ is deduced from the measurement of the recoil velocity $v_r$ of an atom when it absorbs a photon ($v_r=\hbar k/m$ with $\hbar$ the reduced Planck constant, $k$ the wave vector and $m$ the mass of atoms). This measurement is performed by combining a Ramsey-Bord\'e atom interferometer \cite{Borde1989} with the Bloch oscillation technique. Bloch oscillations are used to transfer a large number of recoils to atoms and the interferometer is used to measure the change of velocity induced by the Bloch oscillations. 

\begin{figure}
\includegraphics[width=\linewidth]{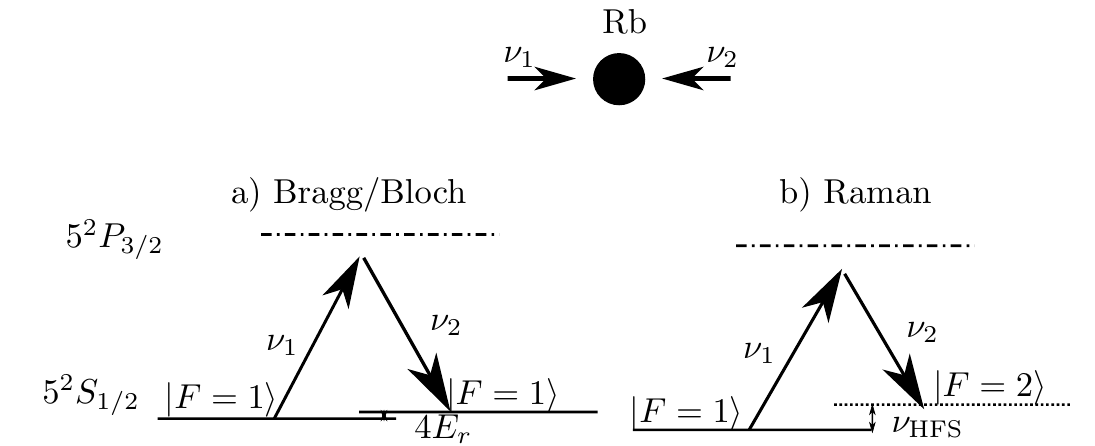}
\caption{\label{fig:blochraman} Atoms are shed with two counter propagating laser beam of frequency $\nu_1$ and $\nu_2$. When they absorb and emit a photon, they acquire twice the recoil velocity. a) In the Bloch oscillation configuration (which is also the same for Bragg diffraction), $\nu_1 \simeq \nu_2$, the transition occurs between the same internal state. The difference between $\nu_1$ and $\nu_2$ accounts only for the Doppler shift induced by the recoil. b) In the Raman configuration, $\nu_1 - \nu_2$ is almost equal to the hyperfine splitting $\nu_{HFS}$, and the transition change the internal state. }
\end{figure}

\begin{figure}
\includegraphics[width=.95\linewidth]{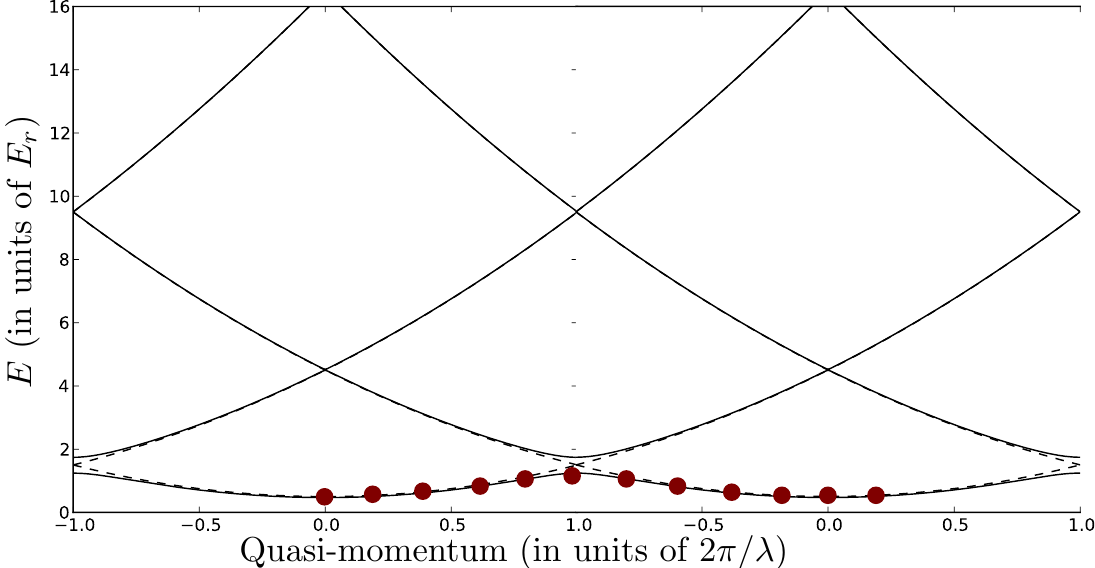}
\caption{\label{fig:bloch_oscilations} Visualisation of a Bloch oscillation. The atoms are
placed in a periodic potential and therefore present a band structure. Eigenstates can be described by the band index and the quasimomentum (x-axis). When an atom prepared in the first band is accelerated, its quasimomentum increases linearly. At the edge of the Brillouin zone (centre of the graph) the atom will follow adiabatically the fundamental band if the acceleration is not too strong. After one oscillation, the atom is back to the initial state. }
\end{figure}

Bloch oscillations (BO) have been first observed in atomic
physics by the groups of Salomon and Raizen \cite{BenDahan,Peik,Wilkinson1996}. The
atoms are shed with two counter-propagating laser beams whose frequency difference
is swept linearly. 
One perspective is to consider that 
the atoms undergo a succession
of  transitions which correspond to the absorption of
one photon from one beam and a stimulated emission of another
photon to the other beam  (see fig.~\ref{fig:blochraman}). The internal state is unchanged
while the atomic velocity increases by 2$\times v_r$ per oscillation.
The Doppler shift due to this velocity variation is periodically
compensated by the frequency sweep and the atoms
are accelerated. A second perspective is to consider that the atoms are
placed in an optical lattice generated by a standing wave, which is accelerated when the
frequency difference between the two laser beams is swept.
The optical lattice induces a periodic optical potential which leads to a band structure; when accelerated it induces an inertial force.
This system is therefore analogous to the Bloch oscillations of an electron in a solid
experiencing an electric field (see figure \ref{fig:bloch_oscilations}). At the edge of the Brillouin zone (center of Fig.~\ref{fig:bloch_oscilations}), where the gap is, atoms are diffracted by the periodic structure (Bragg diffraction). This corresponds to the absorption and stimulated emission of a photon from each laser beam. 

For 87-rubidium atoms, the Doppler shift induced by a variation of velocity of 2$v_r$ is 30 kHz, while the number of Bloch oscillations performed by the atoms is set precisely by the frequency sweep. In our previous work we demonstrated that BO is a very efficient process in terms of photon momentum transfer: more that 500 BOs (corresponding to  1000 recoils) are transferred with less than 50\% of losses \cite{Battesti:04, PhysRevA.95.063604}.

\begin{figure}
\includegraphics[width=.95\linewidth]{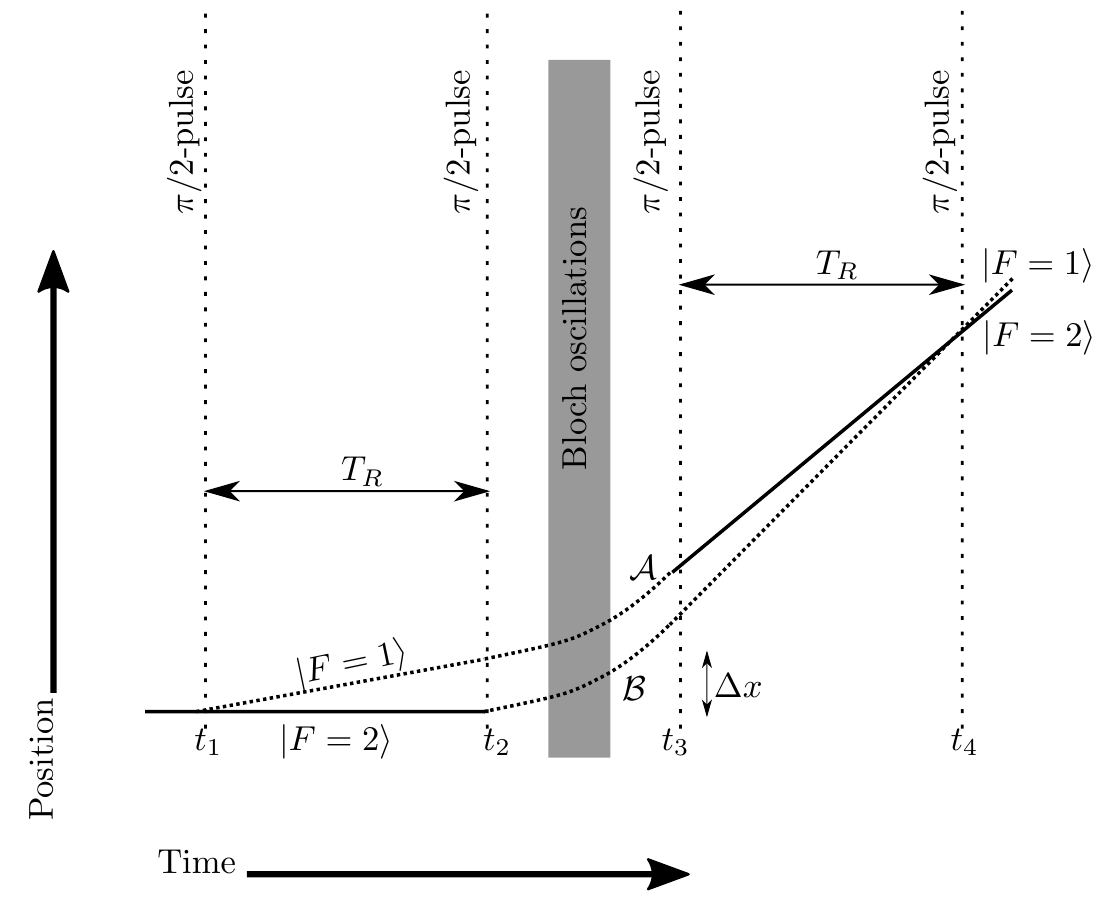}
\caption{\label{fig:schema_interfero} Schematic of the Ramsey-Bordé interferometer used to measure the recoil velocity. A sequence of four Raman $\pi/2$ pulses splits and recombines the atomic wavepacket in two paths (label $\mathcal A$ and $\mathcal B$). Between the second and third pulse, Bloch oscillations are used to accelerate the atoms. }
\end{figure}

The Ramsey-Bordé interferometer is described on Fig.~\ref{fig:schema_interfero}. The atomic wavepackets are manipulated using counterpropagating Raman transition. In our experiment based of $^{87}\mathrm{Rb}$, the Raman transition transfers atoms between the $\left|F=1\right>$ and $\left|F=2\right>$. We denote by $\delta$ the frequency difference between the two lasers. The two laser are counterpropagating: the absorption and stimulated emission of two photons is accompanied by the transfer of two recoils to the atoms. A $\pi/2$ pulse splits a wave packet in two wave packets with different internal and external state. Such a transition also presents a Doppler effect, which is equal to $2kv$ for an atom moving at a velocity $v$ . 

The Ramsey-Bordé interferometer consists of four $\pi/2$ pulses. On Fig.~\ref{fig:schema_interfero} we have drawn the two paths (label $\mathcal{A}$ and $\mathcal{B}$) that interfere. At the output, we observe the fraction of atoms in each state. This fraction depends on the phase shift accumulated by the atoms, and this phase depends on the phase (or frequency) of the laser and of the velocity change induced by BOs.
The phase shift can be calculated using different techniques \cite{PSCT1994, Borde2001, Kleinert2015}. One way consists in integrating the kinetic energy:
\begin{eqnarray}
\Delta\Phi_{\mathrm{at}} & = & \frac 1\hbar\int_{t_1}^{t_4} \frac 12 m v_{\mathcal{A}}^2(t) dt - \frac 1\hbar\int_{t_1}^{t_4} \frac 12 m v_{\mathcal{B}}^2(t) \\
 & = & \frac{m}{\hbar}\int_{t_1}^{t_4} \left( v_{\mathcal{B}}(t) - v_{\mathcal{A}(t)} \right)\left(\frac{v_{\mathcal{B}}(t) + v_{\mathcal{A}}(t)}{2} \right) dt
\end{eqnarray}
Between $t_1$ and $t_2$, $v_{\mathcal{B}}(t) - v_{\mathcal{A}}(t) = -2v_r$; between $t_2$ and $t_3$, $v_{\mathcal{B}}(t)=v_{\mathcal{A}}(t)$ and between, $t_3$ and $t_4$, $v_{\mathcal{B}}(t) - v_{\mathcal{A}}(t) = 2v_r$. This equation gives:
\begin{equation}
\Delta\Phi_{\mathrm{at}} = \frac m\hbar 2v_rT_R\Delta v = 2kT_R\Delta v
\label{eq:sensibility}
\end{equation}
where $T_R = t_2 - t_1 = t_4 - t_3$ and $\Delta v$ is the velocity change induced by the BOs. In this configuration, we use the interferometer as a velocity sensor. Its sensitivity ($\Delta\phi/\Delta v$) is proportional to the separation $\Delta x = 2 v_r T_R$ between the two paths of the interferometer. This can be related to the Heisenberg principle: in order to have a good sensitivity for the measurement of the velocity, one needs to have a large uncertainty in the position, i.e. a large distance between the two paths that the atoms follow simultaneously. 

In order to operate the atom interferometer, we add an additional phase shift by changing the frequency of the laser between pulse 2 and pulse 3. We call $\delta_\mathrm{sel}$ (resp. $\delta_\mathrm{meas}$) the frequency of the laser during the first two pulses (resp. the last two pulses). This will induce an additional phase shift: 
\begin{equation}
\Delta\Phi_{\mathrm{las}} = (\delta_\mathrm{sel} - \delta_\mathrm{meas})T_R
\end{equation}
The total phase shift is given by:
\begin{equation}
\Delta\Phi = \Delta\Phi_{\mathrm{las}} + \Delta\Phi_{\mathrm{at}} = \left(2k\Delta v + \delta_\mathrm{sel} - \delta_\mathrm{meas}\right)T_R
\label{eq:phase_totale}
\end{equation}

In the experiment, we scan the value of $\delta_\mathrm{meas}$ and look for the central fringe, i.e. the value of $\delta_\mathrm{meas}$ such that $\Delta\Phi = 0$. The optimal value corresponds to a change of frequency $\delta_\mathrm{sel} - \delta_\mathrm{meas}$ that compensates the Doppler effect $2k\Delta v$ induced by the Bloch oscillation.

\subsection{Experimental setup}

\begin{figure}
\includegraphics[width=\linewidth]{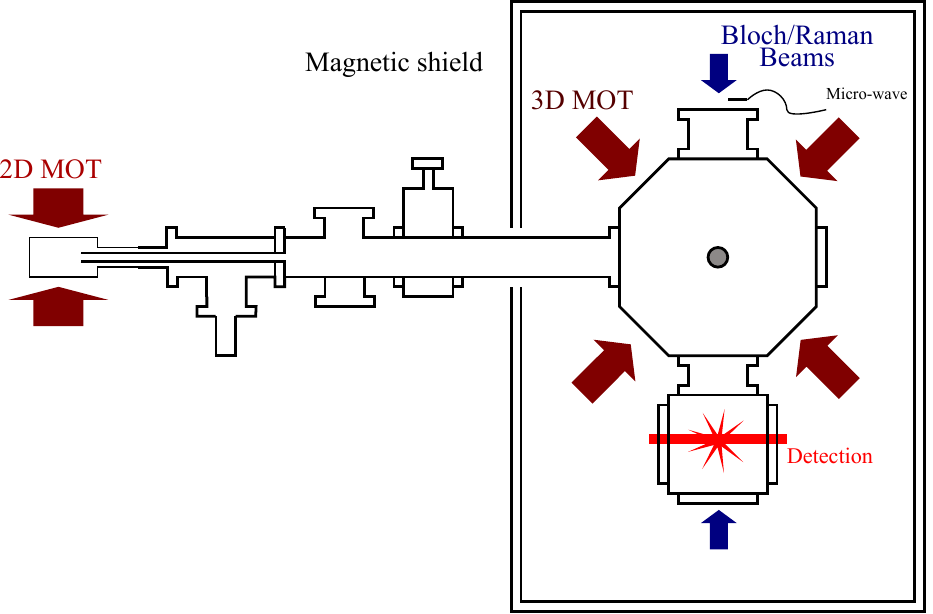}
  \caption{\label{fig:schema_manip} Schematic of the experimental setup used to measure $h/m_\mathrm{Rb}$.
}
\end{figure}

\begin{figure}
\includegraphics[width=\columnwidth]{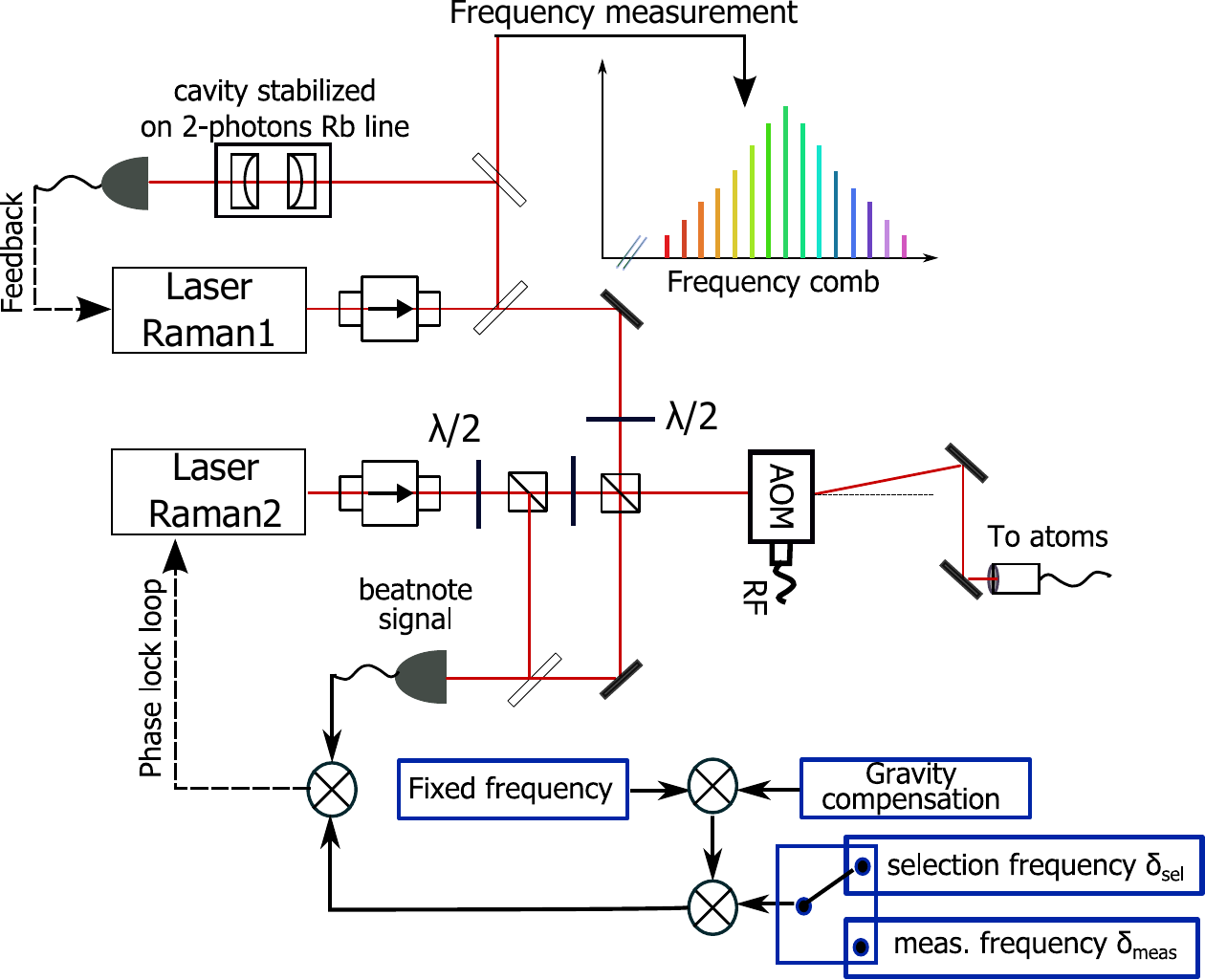}%
  \caption{\label{laser_raman}
   The optical setup of the Raman beam used to perform the atomic interferometer. The two laser diodes are stabilized using an interference-filter-stabilized extended cavity. They are phase-locked. The frequency of one Raman laser is stabilized on an ultra-stable cavity and measured with a femtosecond comb.}
\end{figure}

The experimental set-up is shown on Fig~\ref{fig:schema_manip} and~\ref{laser_raman}. A two-dimensional magneto-optical trap (2D-MOT) produces a
slow atomic beam (about 10$^9$ atoms/s at a velocity of
20 m/s) which loads during 250 ms a three-dimensional
magneto-optical trap. Then a $\sigma^+$-$\sigma^-$ molasses generates
a cloud of about 2 $\times$ 10$^8$ atoms in the $F =2$ hyperfine
level, with a 1.7 mm radius and at a temperature of 4 $\mathrm{\mu K}$. The 2D-MOT cell is a glass cell separated from a UHV-chamber by a differential pumping tube which is also the aperture for the output slow beam. The cooling and pumping lasers are interference-filter-stabilized external-cavity diode lasers (IF-ECL) \cite{Baillard2006}, both lasers are amplified in the same tapered amplifier. 
The Raman lasers are also IF-ECL diode lasers. 
The two diode lasers are phase-locked using a synthesized frequency referenced to a caesium atomic clock. The synthesized frequency results from a mixing of a fixed frequency (6.84 GHz), a frequency ramp  to compensate the fall of atoms in the gravity field (25 MHz/s) and the frequency hopping between $\delta_\mathrm{sel}$ and $\delta_\mathrm{meas}$. For the measurement of $h/m_\mathrm{Rb}$ published in 2010, this was performed by a combination of many synthesizers. This is now conveniently replaced by a single synthesizer based on a field-programmable gate array (FPGA) that produced both the frequency ramp, the frequency hopping as well as an additional phase shift if required \cite{andia:tel-01232238}. 

The Bloch beams originate from a high power laser. The 2.5 W Ti:sapphire laser used in the measurement of 2010 is now replaced by a frequency doubled 1.5~$\mathrm\mu$m fibre laser. This laser produce up to 10~W at 780~nm~ \cite{Andia:15}. The output laser beam is split into two paths, each of which passes
through an AOM to adjust the frequency offset and amplitude before being injected into a polarization maintaining fibre. The depth of the generated optical lattice is 45$E_r$ ($E_r$ is the recoil energy) for an effective power of 150~mW seen by the atoms. The optical scheme of the Bloch and the Raman beams is described in detail in \cite{Bouchendira2011,Cadoret2008}. 

The frequencies of one Raman laser and the Bloch oscillation laser are stabilized onto the same ultra-stable Zerodur Fabry-Perot cavity, itself stabilized on the $5S_{1/2}(F=3)\longmapsto 5D_{3/2} (F=5)$ two-photon transition of 85-rubidium~ \cite{Touahri} (short term). On the long term, these frequencies are precisely measured by using a femtosecond  comb  referenced to the cesium clock. As the measurement of the ratio $h/m_{\mathrm{Rb}}$ is performed in terms of frequency, it is thus directly connected to the cesium standard.

\begin{figure}
  \includegraphics[width=\columnwidth]{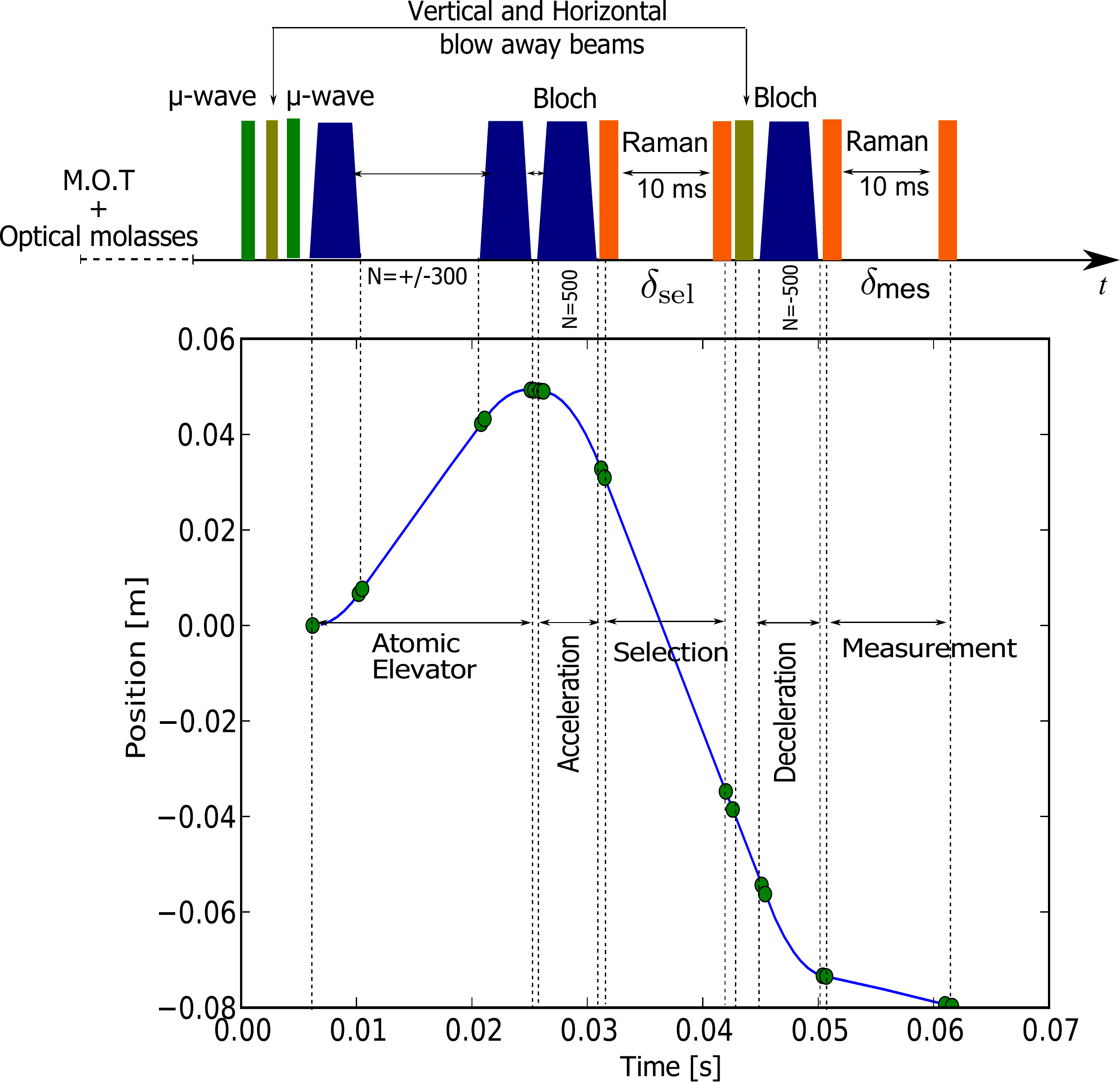}%
  \caption{\label{sequence_complete}
     The pulses timing sequence and atomic trajectory during the measurement procedure.}
\end{figure}

The Fig.~\ref{sequence_complete} shows the precise timing sequence used in the interferometer as well as the trajectory of atoms.  The Bloch oscillations transfer 1000 recoils to the atoms, i.e. a total velocity of about 6~m/s. This leads to a large motion of the cloud (about 10 cm), a motion that is much larger than the distance between the two paths, about 600~$\mathrm{\mu m}$ and that we cannot distinguish on the graph. Furthermore, we prefer to accelerate the atoms before the interferometer and slow them down during the measurement. With this method atoms end up with a smaller velocity and can be efficiently detected. 

Two important effects  were neglected in the simplified calculation of the phase shift in eq.\ref{eq:phase_totale}. The first one is the acceleration of atoms due to gravity, this will add an additional contribution that equal to $\Delta\Phi_{\mathrm{grav}}=2kg(t_3-t_1)T_R$. The second one is the internal energy of atoms which vary due to the second order Zeeman effect (we use $m_F=0$ state) or light shifts. 

To cancel the effect due to the gravity, the atoms are accelerated alternatively upward and downward and the difference between the results eliminate $\Delta\Phi_{\mathrm{grav}}$. Moreover, for each initial acceleration, two spectra are recorded by exchanging the directions of the Raman beams. This changes the relative sign between the Doppler effect and the level shifts, which can then be eliminated. 

In order to operate the interferometer in the same region when we alternate the BO acceleration, we displace the cloud with a set of two Bloch oscillations sequences. In order to prepare the $m_F=0$ state, we use a micro-wave transition to select those atoms in the initial cloud (see Fig.~\ref{sequence_complete}).

\begin{figure}
\begin{center}
\includegraphics[width=.95\linewidth]{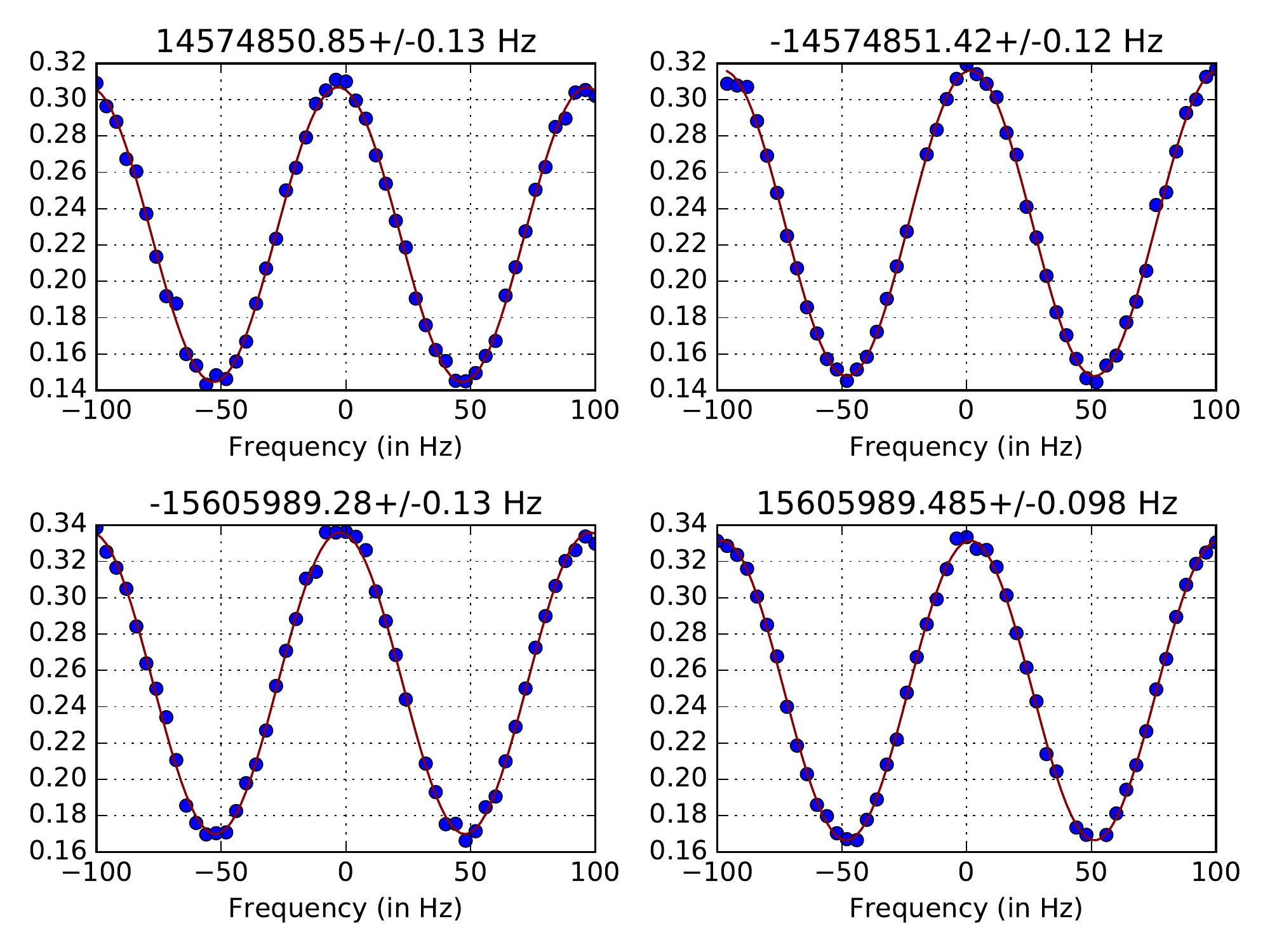}
\end{center}
\caption{\label{fig:spectre} Typical set of four spectra used for the determination of $h/m_\mathrm{Rb}$. The quantity $N_2/(N_1 + N_2)$, where $N_1$ and $N_2$ represent respectively the populations in hyperfine $F=1$ and $F=2$ state, is plotted versus the frequency difference $\delta_\mathrm{sel} - \delta_\mathrm{meas}$. The four spectra are recorded with 50 points each during 4 minutes in total. The measured position of the central fringe is indicated over each spectrum.} 
\end{figure}

A typical fringe pattern is shown in Fig.~\ref{fig:spectre}. The periodicity of the fringe is given by $1/T_R = 100~\mathrm{Hz}$ in this experiment. It is recorded during 4 min where we take 200 points (50 points per spectra). The central fringe in determined with an uncertainty of $0.12~\mathrm{Hz}$. This correspond to a relative uncertainty of $10^{-8}$ on the Doppler shift induced by 500 BOs (about 15~MHz). In comparison to the results obtained in 2010 \cite{Bouchendira2011}, the statistics has been improved: now we obtain a similar uncertainty with half the number of points. This improvement is due to a reduction of the vibrations thanks to a better isolation platform. 


Different techniques could be used to further reduce the impact of the vibrations. It is possible to use an external sensor to measure vibrations and compensate the noise. This technique is used for gravimeter based on atom interferometry (see for example \cite{Merlet2009}). Another possibility consists in running both upward and downward spectra at the same time and extract the phase difference. This technique, similar to the one used in gradiometers \cite{Fixler2007}, was used in the 2018 measurement of Berkeley \cite{Parker2018}.


A value of $h/m_\mathrm{Rb}$ is obtained by recording four spectra using the following:
\begin{equation}
\frac{\hbar}{m_{\mathrm{Rb}}}=\frac{1}{4}\sum_{Spectra}\frac{2\pi\left\vert \delta_\mathrm{sel}-\delta_\mathrm{meas}\right\vert }{2 N k_\mathrm{B}(k_1+k_2)}
\end{equation}
where $k_1$ and $k_2$ are the wave-vectors of the Raman laser beam, $k_\mathrm{B}$ is the wave-vector of the Bloch laser beams and $N$ the number of Bloch oscillations. A set of four spectra, which takes about 4 minutes to acquire, provides a measurement of $h/m_{\mathrm{Rb}}$ with a relative uncertainty of 5$\times$ 10$^{-9}$ (2.5$\times$ 10$^{-9}$ on $\alpha$). 

\begin{figure}
\includegraphics[width=\columnwidth]{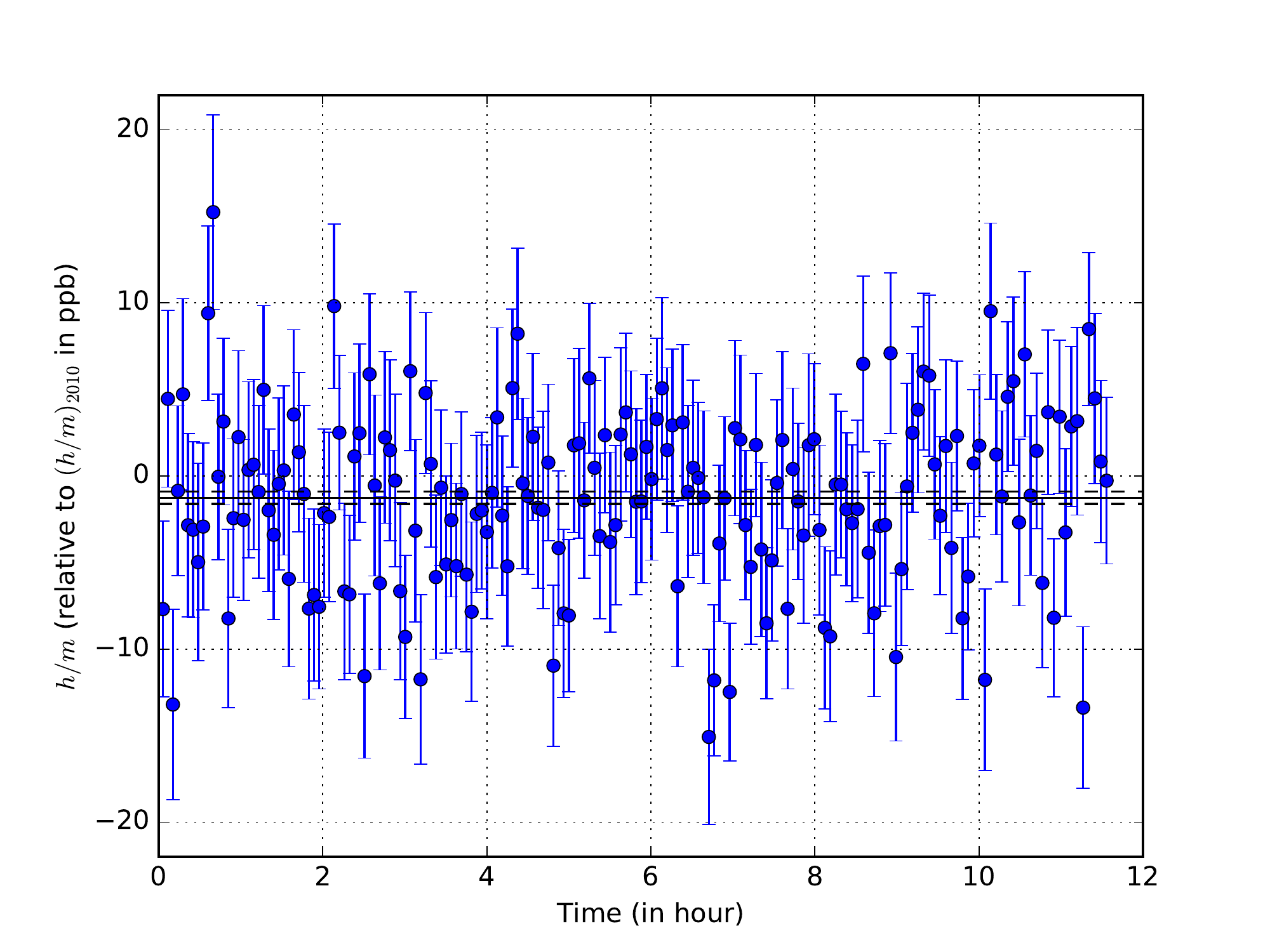}
  \caption{\label{many_points_manu}
A set of 180 measurements of the ratio $h/m_{\mathrm{Rb}}$ obtained during 12 hours of integration time. Raw data are plotted relative to the value of $(h/m)_{2010} =  4.591\ 359\ 2729\ (57) \times 10^{-9}\ \mathrm{m^2 s^{-1}} $. No correction of systematic effects was applied to this set of data. }
\end{figure}

Figure~\ref{many_points_manu} shows a set of 180 determinations of the ratio $h/m_{\mathrm{Rb}}$ recorded during 12 hours. The standard deviation on the mean value is 3.6$\times$10$^{-10}$, with a $ \chi^2/(n-1)$=1.05. These points are among the numerous measurements that have been made since 2010 \cite{andia:tel-01232238}. As we have seen above, statistics was improved. The reliability of the experimental set-up was also improved. However we are not able to overcome some systematic effects and improve the measurement of 2010. 

Before explaining in details the systematic effects that limit the accuracy of the experiment, we can briefly describe possible improvements of the sensitivity. As explained by equation \ref{eq:sensibility}, the sensitivity is proportional to the distance between the two path during the BOs. For the usual Ramsey-Bordé interferometer based on Raman transitions, this distance is proportional to $2v_rT_R$, where $T_R$ is the duration between the first and second $\pi/2$ pulses. This separation can of course be extended by increasing $T_R$ - however the time is limited by physical constraints. In the $h/m$ experiment, this is mainly the displacement of atoms along the propagation axis (atoms are moving at $6~\mathrm{m/s}$). A promising method consist in increasing the velocity splitting of the beam splitter using a so called large momentum transfer (LMT) beam splitter. Different approaches have been investigated using Bloch oscillations \cite{Denschlag, clade_PRL2009, muller2009_PRL240403} or high order Bragg diffraction. This last method is used at Berkeley (with a $10~\hbar k$ beamsplitter \cite{PhysRevLett.115.083002, Parker2018}). It can produce momentum splitting of more than $100~\hbar k$~ \cite{Chiow2011}.

\section{Systematic effects}

\begin{table}
\caption{\label{BudgetError} Error budget on the determination of
$1/ \alpha$ (systematic effect and relative uncertainty in part per $10^{10}$.)}
\centering
\begin{tabular}{|l|r|}
\hline
 &$\delta\alpha/\alpha$~($\mathrm{in}~10^{-10}$)\\
\hline
Laser frequencies& $\pm1.3$\\
Beams alignment&$-3.3\pm 3.3$\\
Wave front curvature and Gouy phase&$-25.1 \pm 3.0$\\
2nd order Zeeman effect&$4.0 \pm 3.0 $\\
Gravity gradient&$-2.0\pm 0.2$ \\
Light shift (one photon transition)&$\pm 0.1$\\
Light shift (two photon transition)& $\pm0.01$ \\
Light shift (Bloch oscillation)& $\pm0.5$ \\
Index of refraction atomic cloud & \\
and atom interactions& $\pm 2.0$ \\ \hline
Global systematic effects&$-26.4\pm 5.9$\\ \hline 
Statistical uncertainty& $\pm 2.0$ \\
Rydberg constant and mass ratio \cite{CODATA2014} & $\pm 1.4$ \\ \hline \hline
Total uncertainty& $\pm 6.2$ \\
\hline
\end{tabular}

\end{table}

The systematic effects that we evaluate for the measurement of 2010 are described in the table~\ref{BudgetError}. The reader will find a precise description of those effects in previous publications \cite{clade:052109, Bouchendira2011}. In the following we will describe them briefly. We will then focus on the wave front curvature and Gouy phase shift, which is the most fundamental limit of the experiment. A large part of our work over the past years was dedicated to this effect. 

The experimental protocol allows to cancel a large part of the level shifts (Zeeman and light shift). This cancellation is performed in three ways: between the selection and the measurement pulses of the interferometer, between the upward and downward trajectories, and when the Raman beams direction is changed. The vacuum chamber is enclosed in a double magnetic shield and we have precisely evaluated the residual magnetic field along the interaction area using Zeeman sensitive Raman transitions. The correction on $\alpha$ due to the second order Zeeman shift is estimated to 4$\times$10$^{-10}$. The light shift is mainly due to the expansion of the atomic cloud between the selection step and the measurement step, and to the unbalance of the laser intensity when we exchange the direction of the Raman beams.  

When we switch the direction of the acceleration, the average position of the two trajectories differs. As gravity is not uniform, a correction proportional to the gravity gradient should be applied. 

The density of the atomic cloud is about 2$\times$10$^8$ atoms/cm$^3$. The refraction index is less than $10^{-10}$ and the phase shift due to mean-field effects is estimated to be less that 10$^{-10}$ on each path. We have taken a conservative uncertainty of 2$\times$10$^{-10}$. Theoretical works indicates that actually no correction of the momentum should be applied to our experiment \cite{Clade2006, Barnett2010a}. Concerning the effect of the mean field effect, we have made a precise model to calculate how the phase shifts compensate between the two path of the interferometer \cite{PhysRevA.92.013616}. This effect, which is relevant for high sensitive experiments based on a Bose-Einstein condensate, will be tested on a new experimental set-up that we have built. 

The main systematic effect comes from the geometry of the laser beam used for the interferometer and the Bloch oscillation. The formula used for the recoil induced by one Bloch oscillation ($2\hbar k/m$) or the Doppler effect of the Raman transition ($(k_1 + k_2) v$) is valid only for perfectly counter propagating plane waves. A first correction comes from the alignment of the different beams (any small angle will induce a negative effect). A second correction comes from the fact that we don't use plane waves but beams that are Gaussian. The effective wave-vector is then given by the gradient of the phase along the measurement axis z. For a Gaussian beam we have calculated that: 
\begin{equation}
k_{\mathrm{eff}}=\frac{d\phi}{dz}=k-\frac{2}{k}\left[ \frac{1}{w^2}-\frac{r^2}{w^4}+\frac{k^2r^2}{4R^2}\right]  
\label{eq:gouy}
\end{equation}
where $r$ is the radius of the atomic cloud, $w$ the waist of the laser and $R$, the curvature radius. The correction is  about $2.5\times 10^{-9}$ on $\alpha$. This is the largest correction. It scales as $1/w^2$. In our 2010 measurement, the waist of the laser was 3.6~mm. A large part of the experimental work conducted since then was to run the experiment with a larger waist, which implies to use more powerful lasers \cite{andia:tel-01232238}. However, as we will see in the next section, this effect is more important than we initially thought. 


\section{Photon momentum and atom recoil}

As explained above, the fundamental process in the determination of the $h/m_\mathrm{Rb}$ is the absorption or stimulated emission of photons by atoms. This process transfers momentum between light and matter. It causes the recoil of atoms. The recoil makes it possible to separate the two paths of the interferometer. In the Raman transitions, the Doppler effect is proportional to the recoil. In the Bloch oscillations, it is used to accelerate the atoms. 

In quantum mechanics, the momentum is well defined for a plane wave only. This applies to atoms as well as it does to light. 
It is easy to be convinced that as far as the atoms are concerned, it is not important if they are not in a plane wave. Indeed, any wave packet can be decomposed into plane waves. As calculated in equation~\ref{eq:sensibility}, the phase shift in the interferometer is independent of the initial velocity, and therefore calculations are correct for a wave packet.  This property is valid for any closed interferometer. 

The reasoning cannot be applied for the light: the phase shift of the interferometer strongly depends on the photon momentum. In the experiment one can use a semi-classical approach for the interaction between light and matter (atoms are quantized but not light). The atom recoil is then given by the phase gradient of the laser beam at the position of the atom. 

In the previous section, we have calculated this term for a Gaussian beam and obtained equation \ref{eq:gouy}.
At the focus of a laser beam, the phase is shifted by $\pi$, this is the Gouy phase. This effect corresponds to the term of equation that scale as $1/w^2$. Even at the waist of a laser beam, where the wavefront is flat and at the centre of the beam, the momentum is reduced by a factor $\frac {2}{kw^2}$. This correction can be interpreted with the dispersion in momentum of the plane wave decomposition of the laser beam. Each plane wave tilted by an angle $\theta$ with respect to the $z$-axis gives a negative correction $-\theta^2/2$. For a Gaussian beam, the divergence is proportional to $\lambda/w$. This reasoning gives the correct order of magnitude for the correction, but is not able to give the correction calculated in \ref{eq:gouy}, which depends on the position $r$ within the beam. 

The formula \ref{eq:gouy} can be generalized. Let us consider a laser beam propagating along the $z$ axis of which we know the amplitude $A(x, y, z_0)$ and phase $\phi(x, y, z_0)$ on the plane $z=z_0$. One can use the Helmholtz equation to propagate the wavefront along the $z$ axis and therefore deduce the $z$ component $k_z$ of the phase gradient at the position $z_0$. We obtain that $k_z = k (1+\delta k_\mathrm{rel})$ with 
\begin{equation}
\label{eq:momentum_correction}
\delta k_{\mathrm{rel}} = -  \frac12\left|\left|\frac{\vec\nabla_{\perp}\phi}{k}\right|\right|^2+\frac1{2k^2}\frac{\Delta_{\perp} A}{A}
\end{equation}
where $\Delta_{\perp}$ and $\vec{\nabla}_{\perp}$ are Laplacian and gradient evaluated in the plane $z=z_0$.

The first term related to the phase gradient corresponds to a tilt in the propagation direction with respect to the $z$-axis due to a local distortion of the wavefront. The second term gives a correction to the momentum even in the case of a plane wavefront. This counter intuitive term is the generalization of the Gouy phase shift. In the following, we will see how to apply this formula in the case of optical aberration and in the case of short scale wavefront distortions. 

A common optical instrument used to characterize wavefronts and the optical aberration is the Shack-Hartmann wavefront sensor. It consists of an array of lenses that focus the beam on a CCD. The local tilt $\vec\nabla_{\perp}\phi$ of the wavefront is then measured by looking precisely to the position of the focal spot of each lens on the sensor. Usually, the global wavefront $\phi$ is reconstructed from $\vec\nabla_{\perp}\phi$. In our situation, we do not need this reconstruction and together with a measurement of the intensity of the spot, one can evaluate eq.~\ref{eq:momentum_correction}. The typical resolution (distance between two lenses) of this instrument is 200~$\mathrm{\mu m}$. It therefore allows to calculate precisely the correction at the expected position of the cloud. 

However, the resolution of the sensor is not good enough to see fluctuations at shorter scale. One would think that those fluctuations would average over the size of the cloud. However we have shown recently that they may cause a systematic effect \cite{PhysRevLett.121.073603}. In this reference, we have studied a random amplitude and phase noise with a given typical size $l$ and relative amplitude $\sigma$. In eq.\ref{eq:momentum_correction}, the first term scales as $\sigma^2/(kl)^2$ while the second one scales as $\sigma/(kl)^2$. Therefore, locally, the dominant contribution comes from intensity variation and not wavefront distortion. Using reasonable values ($l=100~\mathrm{\mu m}$ and $\sigma=5\%$), the amplitude of fluctuations is of the order of $7\times 10^{-8}$, which is much larger than the accuracy of the recoil measurement.

This effect cannot be observed directly, because, in typical experiments, the size of the cloud is much larger than $l$ and the effect is averaged. However, this average is biased, since the fluctuations of the momentum (dominated by $\frac{\Delta_{\perp} A}{A}$) are correlated with the intensity of the laser (local maxima of intensity correspond to negative Laplacian and vice versa). In our experiment, we tend to favor atoms in relatively high intensity zones, because adiabatic following of the fundamental band during Bloch oscillations is better fulfilled, and the momentum transfer is more efficient. Therefore, on average, atoms will see a negative correction to the local momentum. 

\begin{figure}
\includegraphics[width=.95\linewidth]{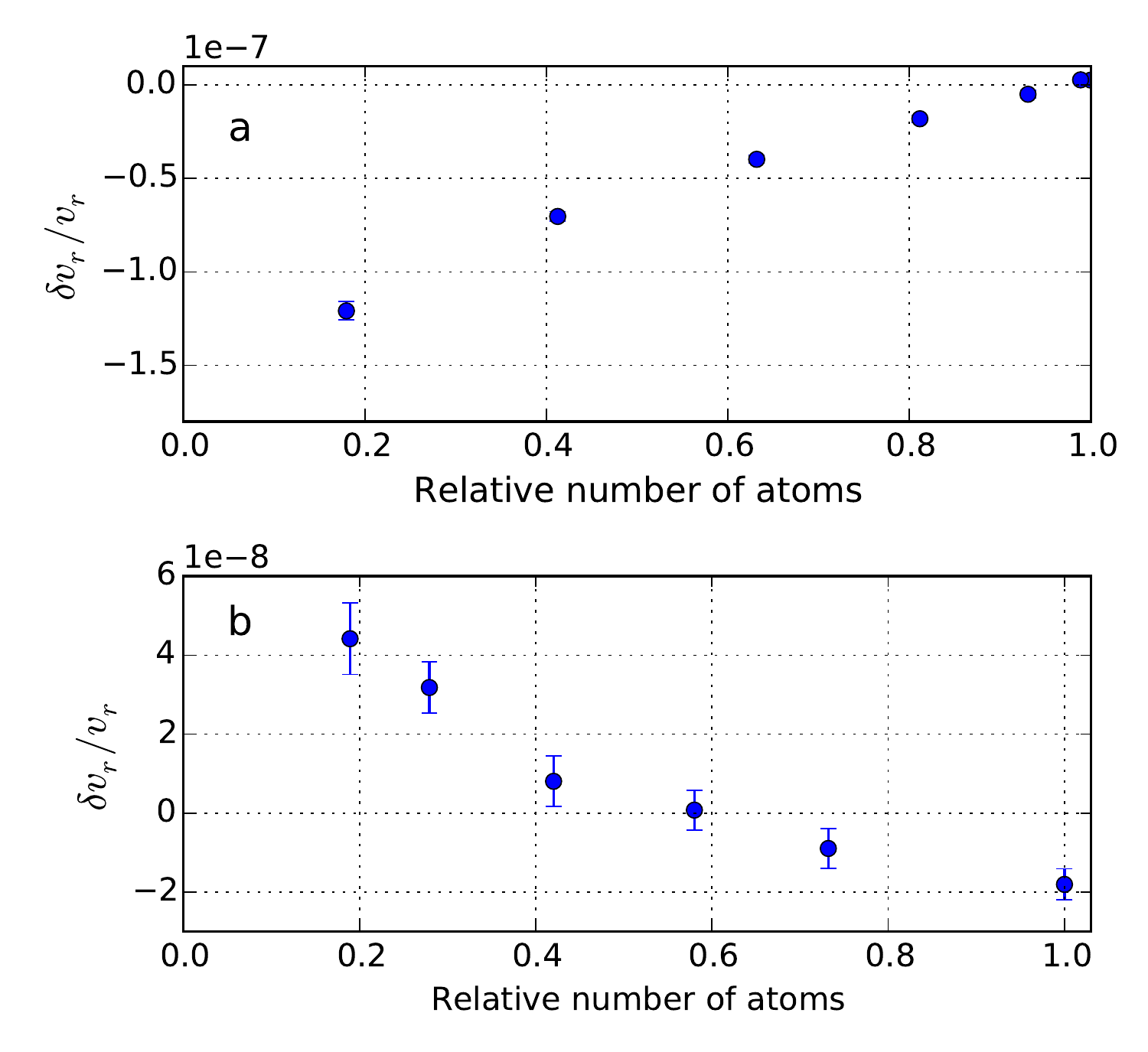}
\caption{\label{fig:effet_systematic} Systematic effect in the measurement of the atomic recoil due to noise in the optical field. We have plotted the measurement of the atomic recoil when we filter the atoms in a zone of relatively high intensity (plot a) or relatively low intensity (plot b). The x-axis represents the number of remaining atom from right (100\%, no selection) to the left (strong selection).}
\end{figure}

Figure~\ref{fig:effet_systematic}a shows recoil measurement where we have been able to exaggerate this effect. To this end, we have reduced the intensity of the Bloch oscillations beam, in such a way that atoms that see an intensity below the average are eliminated. The systematic shift will then scale as $\sigma/(kl)^2$. We have also been able to invert this effect (Fig.~\ref{fig:effet_systematic}b): we select atoms only in the region where the intensity is above the average. In this situation the recoil observed may be larger than $h\nu /c$. This is a very counter intuitive effect: by shining to the atoms photons that can be described by a superposition of plane waves of absolute momentum $h\nu/c$, one is a able to transfer quanta of recoil larger than $h\nu /c$. 

This systematic effect is now dominant in our experiment and we are studying different ways to minimize it: one is to work with parameters for Bloch oscillations such that the probability is independent on the intensity (which implies basically to work at a probability even closer to 100\%) or to increase the duration of the experimental sequence and take benefits of transverse motion of atoms to reduce the correlation between the position of atoms and the local intensity of the laser beam. We should note that this effect has been analysed in the recent experiment of Berkeley and is negligible (supplementary material of \cite{Parker2018}).

\section{Conclusion}

\begin{figure}
\includegraphics[width=.9\linewidth]{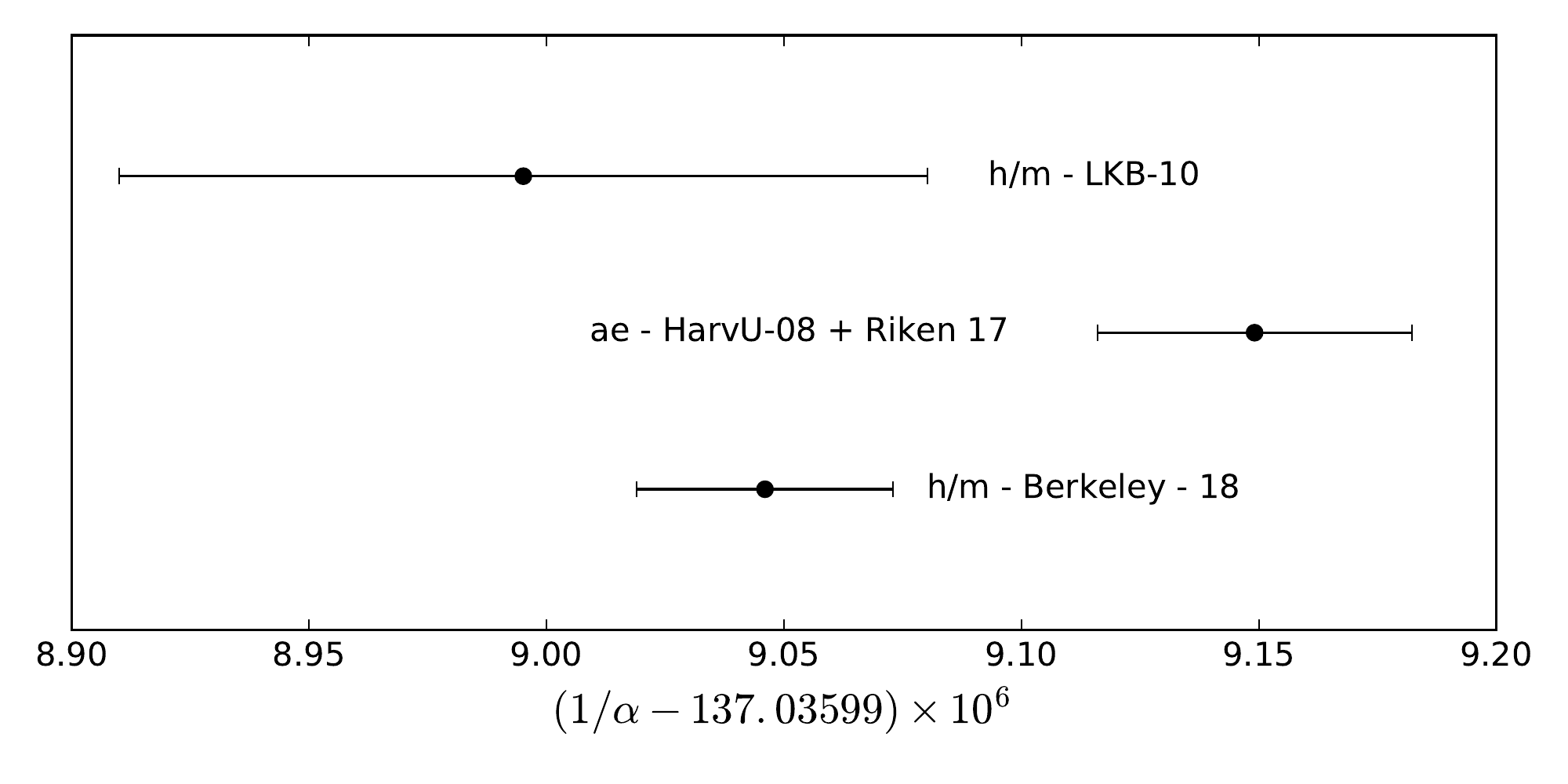}
\caption{\label{fig:comp_alpha_zoom} Zoom on the three best determinations of $\alpha$: the $h/m_\mathrm{Rb}$ measurement of LKB \cite{Bouchendira2011}; the measurement of ${a_\mathrm{e}}$ from Harvard \cite{Hanneke2008} combined with the last calculation from Riken \cite{PhysRevD.97.036001}; and the measurement of $h/m_\mathrm{Cs}$ from Berkeley \cite{Parker2018}.}
\end{figure}

Despite the different improvements made on the experiment since 2010, we are not yet able to present a new measurement due to the systematic effect we have described in the previous section. Our most precise measurement of $h/m_\mathrm{Rb}$ is the one we published in \cite{Bouchendira2011}:
\begin{equation}
\frac h{m_\mathrm{Rb}} =  4.591\ 359\ 2729\ (57) \times 10^{-9}\ \mathrm{m^2 s^{-1}}
\end{equation}

Using the more recent value of $A_r(\mathrm{Rb})$ published by the Atomic Masse Evaluation of 2012 \cite{AMU2012} we obtain:
\begin{equation}
\frac h{m_\mathrm{u}} =  3.990\ 312\ 7193 (49) \times 10^{-9}\ \mathrm{m^2 s^{-1}}
\end{equation}

This leads to the following value of $\alpha$:
\begin{equation}
\alpha^{-1} = 137.035\ 998\ 996\ (85)
\end{equation}
Compared to 2010, the uncertainty is slightly reduced due to an improved measurement of $A_r({m_\mathrm{e}})$ \cite{Sturm2012}. 

On Fig.~\ref{fig:comp_alpha_zoom}, we have plotted the three best available determination of $\alpha$: the measurement of LKB that we have presented in this paper, the measurement using ${a_\mathrm{e}}$ from Harvard \cite{Hanneke2008} combined with the last calculation from Riken \cite{PhysRevD.97.036001} and the recent measurement of $h/m_\mathrm{Cs}$ from Berkeley \cite{Parker2018}. The Berkeley measurement represents a major advance since its accuracy ($2.0\times 10^{-10}$) is slightly better than that of the ${a_\mathrm{e}}$ measurement ($2.4\times 10^{-10}$). It is compatible with the measurement that we made in 2010 but shows a difference of $2.5\sigma$ with the measurement of $\alpha$ using ${a_\mathrm{e}}$.

There are currently several phenomena that physics cannot explain. In particular, some cosmological observations can only be explained by the introduction of a hypothetical form of dark matter and dark energy. High-precision laboratory measurements also lead to phenomena that cannot be explained with the Standard Model: for example, the magnetic moment of the muon is greater than predicted by the Standard Model \cite{PhysRevLett.92.161802} and the radius of the proton measured in muonic hydrogen \cite{Pohl2010} differs from the one measured in hydrogen. Possible explanation is that particles not described by the Standard Model affect the measurements. 
In this context, the measurement of ${a_\mathrm{e}}$ and comparison with calculations play an important role since it can either put constraints on the theory explaining those phenomena or, on the contrary, manifest physics beyond the standard model. 
The current difference of $2.5\sigma$ is not statistically sufficient to conclude that there is physics beyond the standard model and it seems important to make independent measurements with similar or better accuracy to clarify this situation.








\section{Acknowledgments}

The results presented in this paper were supported by: the Agence Nationale pour la recherche
INAQED Project No. ANR-12-JS04-0009, the Cluster of Excellence FIRST-TF and the National Institute of Standards and Technology's Precision Measurement Grant (Grant No. 60NANB16D271).

\bibliography{Pierre}

\end{document}